\documentclass[aps,prd,onecolumn,preprintnumbers,groupedaddress,showpacs,nofootinbib,amssymb]{revtex4}
\usepackage[T1]{fontenc}
\usepackage[latin1]{inputenc}
\usepackage{graphicx}
\usepackage[english]{babel}
\usepackage{amsmath}
\usepackage{amssymb}
\usepackage{amsfonts}
\begin{document}
\def\pp{{\, \mid \hskip -1.5mm =}}
\def\cL{{\cal L}}
\def\be{\begin{equation}}
\def\ee{\end{equation}}
\def\bea{\begin{eqnarray}}
\def\eea{\end{eqnarray}}
\def\tr{{\rm tr}\, }
\def\nn{\nonumber \\}
\def\e{\mathrm{e}}


\title{Dark energy from modified gravity with Lagrange multipliers}

\author{Salvatore Capozziello$^{1,2}$, Jiro Matsumoto$^3$,
Shin'ichi Nojiri$^{3,4}$
and Sergei D. Odintsov$^5$\footnote{Also at Tomsk State Pedagogical
University}}

\affiliation{ $^1$ Dipartimento di Scienze Fisiche Universit`a
``Federico II'' di Napoli and $^2$INFN Sez. di Napoli Compl.
Univ. Monte S. Angelo Ed. N, via Cinthia I- 80126 Napoli, Italy \\
$^3$ Department of Physics, Nagoya University, Nagoya 464-8602, Japan \\
$^4$ Kobayashi-Maskawa Institute for the Origin of Particles and
the Universe, Nagoya University, Nagoya 464-8602, Japan \\
$^5$Instituci\`{o} Catalana de Recerca i Estudis Avan\c{c}ats
(ICREA) and Institut de Ciencies de l'Espai (IEEC-CSIC), Campus
UAB, Facultat de Ciencies, Torre C5-Par-2a pl, E-08193 Bellaterra
(Barcelona), Spain}

\begin{abstract}

We study scalar-tensor theory, k-essence and modified gravity with Lagrange
multiplier constraint which role is to reduce the number of degrees of freedom.
Dark Energy cosmology of different types ($\Lambda$CDM, unified inflation with
DE, smooth non-phantom/phantom transition epoch) is reconstructed in such
models. It is demonstrated that presence of Lagrange multiplier simplifies the
reconstruction scenario. It
is shown that mathematical equivalence between scalar theory and $F(R)$ gravity
is broken due to presence of constraint.
The cosmological evolution  is defined by the
second $F_2(R)$ function dictated by the constraint. The convenient $F(R)$
gravity sector  is relevant for local
tests. This opens the possibility to make originally non-realistic theory
to be viable by adding the corresponding constraint.
A general discussion on the role of Lagrange multipliers to make
higher-derivative gravity canonical is developed.

\end{abstract}

\pacs{95.36.+x, 98.80.Cq}

\maketitle

\section{Introduction}

The understanding of late-time cosmic acceleration (the so-called Dark
Energy (DE) epoch) is one of the main challenges of modern
cosmology. There is no yet totally convincing theoretical DE model which is
related also with the lack of precise observational data.
This requests the introduction of new DEs in order to present fundamental and
simple  DE.
It seems that standard strategy of combination
of known models does not work and some new approach should be
invented. Recently, a quite interesting new DE
model \cite{Lim:2010yk,Gao:2010gj} was proposed which consists of two scalar
fields where one of scalars represents the Lagrange multiplier.
This multiplier puts the constraint of special form on the second
scalar field. As a result, the whole system contains the single
dynamical degree of freedom. It was shown that energy of such
theory flows along time-like geodesics which is similar to the
dust. Nevertheless, the theory contains non-zero pressure (dusty
dark energy) \cite{Lim:2010yk,Gao:2010gj}. It may be very natural candidate
for unification of Dark Energy and Dark Matter. It is interesting to study
the role of such novel construction in the known DE models because it may
completely change the structure of theory, its cosmological solutions and
cosmological perturbations theory and may give better and/or qualitatively
different fit towards to observational data. Moreover, the role of Lagrange
multipliers may be twofold because they are often used for reducing of
higher-derivative
gravity systems in canonical formulation. Hence, the most natural area for the
study of Lagrange multiplier constraint is modified gravity. In addition,
adding such constraint in modified gravity may significally improve the
ultraviolet properties of the covariant theory \cite{nojiri2010} leading 
to its
renormalizability conjecture.

In the present letter we discuss the role of such Lagrange
multiplier constraint in modified gravity. This is the first study of this
sort, so the number of models are discussed in order to understand the general
role of Lagrange multiplier construction and its impact to cosmology.
We start from unified
(phantom/canonical) scalar theory with Lagrange multiplier
constraint. The reconstruction scenario for such theory as well as for
k-essence theory is
developed and shown to be significally simpler than in the case without
constraint. Dark Energy cosmology of different types is
reconstructed including standard $\Lambda$CDM, unified
inflation-Dark Energy era and non-phantom/phantom transition
cosmology which turns out to be smooth.
    The comparison with
standard theory is done, it is shown that different potentials
describe such cosmological evolution. Moreover, convenient scalar
does not propagate and does not generate the extra force. Despite
the fact that theory is described by single degree of freedom, it
may be represented in BD-form where Lagrange multiplier has
ghost-like kinetic term.

The relation of such theory with modified $F(R)$-gravity is
investigated. It is shown that mathematical equivalence between
scalar theory and $F(R)$-gravity is broken because of the presence
of Lagrange multiplier constraint.
New constrained form of modified gravity is introduced. It
consists of standard $F_1(R)$-term and Lagrange multiplier which
constrains kinetic-like term for curvature by second function
$F_2(R)$. The cosmological dynamics of such modified gravity
depends only from the choice of $F_2(R)$ function. Dark energy
cosmology may be easily reconstructed using its form.

Unlike to scalar theory, the Lagrange multiplier propagates.
Hence, the role of $F_1(R)$ is in modification of Newton law which
may cause the extra modification of constraint in order to
reproduce the standard Newton law. This is also novel property, as in
convenient modified gravity if theory does not pass newton law regime it is
ruled out as realistic one.

Another application of Lagrange multiplier in canonical formulation is
discussed.
General method of reduction of higher derivatives is developed. It can be
adopted for any higher-order
theories. Specifically,
gravity theories of order $(2k+4)$, with $k$ an integer, need $(k+1)$ Lagrange
multipliers to be made canonical, as we will discuss below.

The layout of this letter is the following. In Sec.II,
we study scalar-tensor and k-essence theories with Lagrange multiplier 
constraint.
DE cosmologies which are generated by the self-interaction potential and
Lagrange multiplier constrained fluid are investigated.
Sec.III is devoted to the same issue but $F(R)$ gravity is considered.
In this case, cosmological dynamics is qualitatively changed due to lack of
mathematical equivalence with scalar-tensor theory and presence of second
$F_2(R)$ function caused by Lagrange multiplier constraint.
In Sec.IV, general considerations on Lagrange multipliers are developed.
We show that a given gravity theory of order $(2k+4)$ can be made canonical by
introducing $(k+1)$ suitable Lagrange multipliers. Conclusions are drawn in
Sec.V.

\section{Dark energy in scalar theory with Lagrange multiplier}

In this section, we consider the accelerating FRW cosmology in the
theory with two scalars where one of scalars is Lagrange
multiplier which constrains the field equation of second scalar.
The starting action has the following form:
\be
\label{LagS1}
S = \int d^4 x \sqrt{-g} \left\{
\frac{R}{2\kappa^2} - \frac{\omega(\phi)}{2} \partial_\mu \phi
\partial^\mu \phi - V(\phi)
- \lambda \left( \frac{1}{2} \partial_\mu \phi \partial^\mu \phi +
U(\phi) \right) \right\}\, .
\ee
Here $\lambda$ is the Lagrange
multiplier field. Depending on the sign of the potential
$\omega(\phi)$ (see ref.\cite{Nojiri:2005pu}) the first scalar
could be the canonical scalar or the phantom. The Einstein
equation has the following form:
\be
\label{LagE} \frac{1}{2\kappa^2} \left(R_{\mu\nu} - \frac{1}{2}
g_{\mu\nu} R \right) = \frac{1}{2} g_{\mu\nu} \left\{
    - \frac{\omega(\phi)}{2} \partial_\rho \phi \partial^\rho \phi -
V(\phi) - \lambda \left( \frac{1}{2} \partial_\rho \phi
\partial^\rho \phi + U(\phi)
\right) \right\} + \frac{\omega(\phi) + \lambda}{2} \partial_\mu
\phi \partial_\nu \phi \, .
\ee
We now work in the FRW metric with
flat spatial part:
\be
\label{Lag7} ds^2 = - dt^2 + a(t)^2 \sum_{i=1,2,3}
\left(dx^i\right)^2\, .
\ee
Then by the variation over $\lambda$,
we obtain
\be
\label{LagS2} 0 = \frac{{\dot\phi}^2}{2} - U(\phi)\, .
\ee
The FRW equations are given by
\bea
\label{LagS3}
\frac{3}{\kappa^2} H^2 &=& \frac{\omega(\phi) +
\lambda}{2}{\dot\phi}^2 + V(\phi) + \lambda U(\phi) = \left(
\omega(\phi) + 2\lambda \right) U(\phi) + V(\phi) \, ,\nn
\label{LagS4} - \frac{1}{\kappa^2} \left(2 \dot H + 3H^2 \right)
&=& \frac{\omega(\phi) + \lambda}{2}{\dot\phi}^2 - V(\phi) -
\lambda U(\phi) = \omega(\phi) U(\phi) - V(\phi) \, .
\eea
These equations show that the EoS-parameter
$w_{\phi\lambda}$ has the following form:
\be
\label{LagS4b}
w_{\phi\lambda} = \frac{\omega(\phi) U(\phi) -
V(\phi)}{\left( \omega(\phi) + 2\lambda \right) U(\phi) +
V(\phi)}\, .
\ee
Eq.(\ref{LagS2}) can be integrated as
\be
\label{LagS5}
t = \pm \int^\phi
\frac{d\phi}{\sqrt{2U(\phi)}}\, .
\ee
as long as $U(\phi)$ is
positive. Then by solving Eq.(\ref{LagS5}) with respect to $\phi$,
one can find the $t$-dependence of $\phi$: $\phi=\phi_0(t)$.
Substituting the expression of $\phi_0(t)$ into (\ref{LagS4}), we
obtain a differential equation for $H$, which gives the
$t$-dependence of $H$: $H=H_0(t)$. Finally substituting
$\phi_0(t)$ and $H_0(t)$ into (\ref{LagS3}), we can find the
$t$-dependence of the Lagrange multiplier field $\lambda$:
\be
\label{LagS6}
\lambda = -
\frac{\omega\left(\phi_0\left(t\right)\right)}{2} +
\frac{1}{2U\left(\phi_0\left(t\right)\right)} \left\{
\frac{3}{\kappa^2} H_0(t)^2 -
V\left(\phi_0\left(t\right)\right)\right\}\, .
\ee
Note that due
to the constraint (\ref{LagS2}), the dynamics is completely
changed. Usually if the scalar field exists, the propagation of
the field generates an extra force and often violates the
observational constraints on the Newton law. Due to the constraint
(\ref{LagS2}), however, there does not appear the propagating mode
of $\phi$. To see this, we consider the perturbation from the
solution $\phi_0 (t)$ given by (\ref{LagS5}),
\be
\label{LagS7}
\phi = \phi_0 (t) + \delta \phi\, .
\ee
Here we assume that $\delta\phi$ can depend on both of the time
coordinate $t$ and the spatial coordinates. Then the constraint
equation (\ref{LagS2}) gives,
\be
\label{LagS8}
0 = \frac{d\delta\phi}{dt} - \frac{U'(\phi)}{\frac{d\phi_0}{dt}}
\delta\phi \, .
\ee
When $U'(\phi)/\frac{d\phi_0}{dt}>0$, the perturbation
$\delta\phi$ grows with time and therefore the solution $\phi_0$
becomes unstable. On the other hand, when
$U'(\phi)/\frac{d\phi_0}{dt}<0$, the solution $\phi_0$ becomes
stable. Anyway there does not appear the oscillating mode in
$\delta\phi$ and therefore $\delta\phi$ does not propagate nor
does not generate the extra force. We should also note that in the
equations (\ref{LagS2}) and (\ref{LagS4}), there does not appear
the term containing the derivatives of $\lambda$, which shows that
the multiplier field does not propagate. We now consider the
reconstruction, that is, when the behavior of the Hubble rate $H$:
$H=H_0(t)$ is known, how can we construct the action in the form
of (\ref{LagS1}) to reproduce $H_0(t)$. First we choose $U(\phi)$
appropriately so that we can easily integrate Eq. (\ref{LagS5}) to
find explicit form of $\phi_0(t)$ and $t = t(\phi)$. Then Eq.
(\ref{LagS4}) gives
\be
\label{LagS9} V(\phi) = \frac{1}{\kappa^2}
\left(2 \dot H_0 \left(t\left(\phi\right)\right) + 3 H_0
\left(t\left(\phi\right)\right)^2 \right) + \omega(\phi) U(\phi)
\, .
\ee
Here $\omega(\phi)$ can be arbitrary. Hence, one finds
that choosing $V(\phi)$ as (\ref{LagS9}), the arbitrary Hubble
rate $H=H_0(t)$ can be reproduced.

Some examples may illustrate the reconstruction scheme. Just for
simplicity, we assume
\be
\label{LagS10}
\omega(\phi) = 1\, ,\quad U(\phi) =
\frac{m^4}{2}\, .
\ee
Here $m$ is a constant with the dimension of
mass and canonical scalar is considered. Then Eq. (\ref{LagS5})
tells
\be
\label{LagS11}
\phi = m^2 t\, .
\ee
Here we have chosen
the $+$ sign in $\pm$ in (\ref{LagS5}). Then Eq. (\ref{LagS9}) has
the following form:
\be
\label{LagS12}
V(\phi) = \frac{1}{\kappa^2}\left\{ 2 \dot H_0
\left( \frac{\phi}{m^2} \right) + 3 H_0 \left( \frac{\phi}{m^2}
\right)^2 \right\} + \frac{m^4}{2}\, .
\ee
Eq. (\ref{LagS6})
indicates that
\be \label{LagS13} \lambda = -1 - \frac{2}{\kappa^2} \dot H_0
\left( \frac{\phi}{m^2} \right) \, .
\ee
Let us consider $H_0(t)$ corresponding to the $\Lambda$CDM model:
\be
\label{LagS14}
H_0(t) = \frac{2}{3l} \coth \left( \frac{t}{l}
\right)\, .
\ee
Here $l$ is the length parameter related with the
cosmological constant. The potential $V(\phi)$ becomes a constant
\be
\label{LagS15} V(\phi) = \frac{4}{3l^2 \kappa^2} +
\frac{m^4}{2}\, ,
\ee
which may be regarded as a cosmological
constant. The scalar field plays the role of
dust\cite{Lim:2010yk,Gao:2010gj}. Thus, one obtains dark energy produced by
dusty Lagrange multiplier field. Following the proposal of
ref.\cite{Nojiri:2005pu}, one can reconstruct the model unifying
the inflation and the late-time acceleration:
\be
\label{LagS16}
H_0 (t) = \frac{H_I + H_L \frac{t^2}{t_0^2}}{1
+ \frac{t^2}{t_0^2}}\, .
\ee
When $t\to 0$, $H$ behaves as $H =
H_I + \mathcal{O}\left(t^2\right)$. Then the universe is almost de
Sitter space corresponding to the inflation. On the other hand,
when $t\to \infty$, $H$ behaves as $H = H_L +
\mathcal{O}\left(t^{-2}\right)$ and the universe becomes
asymptotically de Sitter space describing dark energy era.
Applying the above procedure one gets:
\be
\label{LagS17}
V(\phi) = \frac{1}{\kappa^2} \frac{ - 4 \left(
H_I - H_L \right) \frac{\phi}{m^2 t_0^2} + 3 \left( H_I + H_L
\frac{\phi^2}{m^4 t_0^2} \right)^2}{\left( 1 + \frac{\phi^2}{m^4
t_0^2} \right) } + \frac{m^4}{2}\, .
\ee
As a final example, we
consider the dark energy model which admits the transition
\cite{Nojiri:2005sx} from non-phantom phase to phantom phase:
\be
\label{LagS18} H_0 (t) = h_0 \left( \frac{1}{t} + \frac{1}{t_s -
t} \right)\, . \ee Then we find \be \label{LagS19} V(\phi) =
\frac{1}{\kappa^2}\left( \frac{m^4 \left(3h_0^2 -
2h_0\right)}{\phi^2} + \frac{ m^4 \left(3h_0^2 -
2h_0\right)}{\left(m^2 t_s - \phi\right)^2} + \frac{ 6 h_0^2 m^4
}{\phi \left(m^2 t_s - \phi\right)^2} \right) + \frac{m^4}{2}\, .
\ee
The theory with above scalar potential describes the
transition from non-phantom phase to phantom era. The EoS
parameter $w_{\phi\lambda}$ (\ref{LagS4b}):
\be
\label{LagS23}
w_{\phi\lambda} = -1 - \frac{2t - t_s}{3h_0 t_s}\, ,
\ee
which
shows the phantom crossing ($w=-1$ crossing) when $t=t_s/2$, that
is, $w>-1$ when $t<t_s/2$ and $w<-1$ when $t>t_s/2$. Eq.
(\ref{LagS18}) demonstrates that there is a Big Rip singularity at
$t=t_s$ (for first works on Big Rip singularity, see \cite{rip}).
To understand better the role of Lagrange multiplier
constraint we consider single scalar-tensor theory
\cite{Nojiri:2005pu,Capozziello:2005tf},
\be
\label{LagS20}
S =
\int d^4 x \sqrt{-g} \left\{ \frac{R}{2\kappa^2}
    - \frac{\omega(\phi)}{2} \partial_\mu \phi \partial^\mu \phi - V(\phi)
\right\}\, .
\ee
The reconstruction can be performed by choosing
\be
\label{k6}
\omega(\phi)=- \frac{2}{\kappa^2}H_0'(\phi)\ ,\quad
V(\phi)=\frac{1}{\kappa^2}\left(3H_0(\phi)^2 + H_0'(\phi)\right)\, .
\ee
Then it follows
\be
\label{LagS21}
\phi=t\, ,\quad H=f(t)\, .
\ee
Especially in case of non-phantom/phantom transition
(\ref{LagS18}), one gets\cite{Nojiri:2005pu}
\be
\label{LagS22}
\omega(\phi) = h_0 \left\{ - \frac{1}{\phi^2} + \frac{1}{\left(
t_s - \phi\right)^2} \right\} \ ,\quad V(\phi) =
\frac{1}{\kappa^2}\left( \frac{ \left(3h_0^2 - h_0\right)}{\phi^2}
+ \frac{ 3h_0^2 - h_0}{\left( t_s - \phi\right)^2} + \frac{ 6
h_0^2 }{\phi \left( t_s - \phi\right)^2} \right) \, .
\ee
The form
of the potential is similar to that in (\ref{LagS19}) but the
coefficients are different from each other. We should also note
that in (\ref{LagS19}), the mass dimension of the scalar field is
unity as standard but that of the scalar field in (\ref{LagS22})
is minus unity by following
\cite{Nojiri:2005pu,Capozziello:2005tf}. The resemblance of two
potentials (\ref{LagS19}) and (\ref{LagS22}) comes from the
expressions (\ref{LagS9}) or more explicitly (\ref{LagS12}) and
(\ref{k6}). Note that in (\ref{LagS12}), the scalar field is
chosen, essentially to be time coordinate as in (\ref{LagS11}) by
the choice of $U(\phi)$ in (\ref{LagS10}). Expression
(\ref{LagS9}) comes from the effective pressure but (\ref{k6})
comes from the difference between the effective pressure and the
effective energy density. Then the coefficients in two potentials
are different from each other. In spite of the similarity in the
two models (\ref{LagS20}) and (\ref{LagS1}), there is a big
difference in them, that is, the scalar in (\ref{LagS20})
propagates and often gives a correction to the Newton law but the
scalars in (\ref{LagS1}) do not propagate and do not give any
correction to the Newton law. Now both of the scalar-tensor
theories with a constraint (\ref{LagS19}) and without a constraint
(\ref{LagS22}) give the identical EoS parameter $w$:
$w=w_{\phi\lambda}$ in (\ref{LagS23}). Then in both of the models,
the phantom crossing occurs when $t=t_s/2$ and there is a Big Rip
singularity at $t=t_s$. We should note that $\omega(\phi)$
vanishes when the phantom crossing occurs at $t=t_s/2$, which
shows that if we redefine the scalar field as
\be
\label{f5}
\varphi=\int d\phi \sqrt{\left|\omega(\phi)\right|}\, ,
\ee
the action (\ref{LagS20}) can be rewritten as
\be
\label{f6}
S=\int
d^4 x \sqrt{-g}\left\{\frac{1}{2\kappa^2}R \mp
\frac{1}{2}\partial_\mu \varphi
\partial^\mu \varphi - \tilde V(\varphi)\right\}\, .
\ee
The sign in front of the kinetic term depends on the sign of
$\omega(\phi)$. Therefore at the point of the phantom transition,
the sign changes. In this sense, in the model (\ref{LagS20}), the
phantom transition is not smooth and it has been shown that the
transition is very unstable
\cite{Nojiri:2005pu,Capozziello:2005tf}. On the other hand, for
the model with a constraint (\ref{LagS1}), the transition seems to
be smooth. Finally in this section, we show that the action
(\ref{LagS1}) can be rewritten as the Brans-Dicke theory coupled
with a (phantom) scalar field. Let $\omega(\phi=1$ and consider
the conformal transformation
\be
\label{LagBD1}
g_{\mu\nu}=
\e^\sigma\, , \quad \e^{\sigma/\sqrt{3}} \equiv \left( 1 + \lambda
\right)^{-1}\, .
\ee
Then the action (\ref{LagS1}) has the
following form:
\be
\label{LagBD2}
S = \int d^4 x \sqrt{-g}
\left\{ \frac{\e^{\sigma/\sqrt{3}}}{2\kappa^2}\left( R +
\frac{1}{2}\partial^\mu \sigma \partial_\mu \sigma \right)
    - \frac{1}{2} \partial_\mu \phi \partial^\mu \phi - \e^{2\sigma/\sqrt{3}}
V(\phi) - \left(\e^{2\sigma/\sqrt{3}} -
\e^{\sigma/\sqrt{3}}\right) U(\phi) \right\}\, .
\ee
Now the
constraint disappears and the scalar field $\phi$ and the
ex-Lagrange multiplier field $\lambda$ or $\sigma$ seems to be
propagating. However, the kinetic term of $\sigma$ is not
canonical (ghost-like). Then effectively the amplitude coming from
the propagation of $\phi$ might be canceled by the propagation of
$\sigma$. One may choose $V(\phi)$ so that the smooth phantom
crossing occurs in our model. Thus, we demonstrated that the
theory with two scalars where one of them is Lagrange multiplier
may lead to variety of dark energy cosmologies. Despite the fact
that the theory possesses single dynamical degree of freedom, its
cosmology seems to be qualitatively different from the one of the
theory with single scalar.

It is well-known that scalar-tensor theory is mathematically
equivalent to $F(R)$ gravity which does not lead to physical
equivalence of two theories\cite{capo}. Now we may try to
transform the action (\ref{LagS1}) into the $F(R)$-gravity form.
Making the conformal transformation
\be
\label{LagBD3} g_{\mu\nu}
\to \e^{\kappa\phi\sqrt{\frac{2}{3}}} g_{\mu\nu} \, ,
\ee
the kinetic term of $\phi$ is canceled and one obtains
\be
\label{LagBD4} S = \int d^4 x \sqrt{-g} \left\{
\frac{\e^{\kappa\phi\sqrt{\frac{2}{3}}} R}{2\kappa^2}
    - \e^{2\kappa\phi\sqrt{\frac{2}{3}}} V(\phi)
    - \lambda \left( \frac{1}{2} \e^{\kappa\phi\sqrt{\frac{2}{3}}} \partial_\mu
\phi \partial^\mu \phi
    - \e^{2\kappa\phi\sqrt{\frac{2}{3}}} U(\phi) \right)
\right\}\, .
\ee
In the standard case, there is no last term in
the above action and scalar becomes the auxiliary one (no the
kinetic term). Since the term containing $\lambda$ includes the
derivative of $\phi$, we cannot integrate and/or delete the scalar
field $\phi$ and therefore it is difficult to rewrite the action
(\ref{LagS1}) in the $F(R)$-gravity form. Hence, the mathematical
equivalence between two theories seems to be broken due to
presence of the Lagrange multiplier term.

As an extension of the
scalar field with constraint, we may consider k-essence model\cite{k} with
a constraint:
\be \label{c1} S= \int d^4 x \sqrt{-g} \bigg \{
\frac{R}{2\kappa^2} + K(\phi, X) + \lambda \Big( X- U(\phi) \Big)
    + L_\mathrm{matter} \bigg \} ,\quad X \equiv - \frac{1}{2}\partial^\mu \phi
\partial_\mu \phi\, ,
\ee
where $K$ is an appropriate function of $\phi$. Let us check if the
introduction of Lagrange multiplier in such theory does not destroy its
consistency and the reconstruction discussed in \cite{Matsumoto:2010uv}. The
FRW equations are given by
\be
\label{c2} \frac{3}{\kappa^2} H^2 = 2 X \frac{\partial K\left(
\phi, X \right)}{\partial X}
    - K\left( \phi, X \right) +2 \lambda X + \rho_\mathrm{matter}(t)\, ,\quad
    - \frac{1}{\kappa^2}\left(2 \dot H + 3 H^2 \right)
= K\left( \phi, X \right) + p_\mathrm{matter}(t)\, .
\ee
Here we include the contribution of matters with a constant EoS
parameters $w_i$. Then the energy density $\rho_\mathrm{matter}$
and the pressure $p_\mathrm{matter}$ of matters are given by
\be
\label{c3} \rho_\mathrm{matter} = \sum_i \rho_{0i}
a^{-3(1+w_i)}\, ,\quad p_\mathrm{matter} = \sum_i w_i\rho_{0i}
a^{-3(1+w_i)}\, .
\ee
Then the variation of (\ref{c1}) with respect to $\lambda$ gives,
\be
X-U(\phi)=0\, . \label{c4}
\ee
We choose $U(\phi)=m^4/2$. So (\ref{c4}) gives $\phi=m^2t$, then
one can rewrite the equations in (\ref{c2}) in the following form
\bea
&& \label{c5a} K\left( m^2 t, m^4/2 \right) =
-\frac{1}{\kappa^2}\left(2 \dot H + 3 H^2 \right)
    - \sum_i w_i\rho_{0i} a^{-3(1+w_i)}\, , \\
&& \label{c5b} \left. \frac{\partial K\left( m^2 t, X
\right)}{\partial X}\right|_{X=m^4/2} = - \frac{2}{\kappa^2} \dot
H -\lambda - \sum_i \left(1+w_i\right) \rho_{0i} a^{-3(1+w_i)}\, .
\eea
Then by using an appropriate function $g(\phi/m^2)$, if we choose
\be
\label{cc1}
K\left( m^2 t, m^4/2 \right) =
-\frac{1}{\kappa^2}\left(2 g''(\phi/m^2) + 3 {g'(\phi/m^2)}^2
\right)
    - \sum_i w_i\rho_{0i} a_0^{-3(1+w_i)} \e^{-3(1+w_i)g(\phi/m^2)}\, ,
\ee
one gets the following solution of (\ref{c5a}):
\be
\label{c7}
H=
g'(t) \quad \left(a = a_0 \e^{g(t)} \right)\, .
\ee
Note that
$X$-dependence of $K(\phi,X)$ can be arbitrary as long as
$K(\phi,X)$ satisfies (\ref{cc1}). The $t$-dependence of
$\lambda(t)$ can be determined by using (\ref{c5b}) as
\be
\label{cc2}
\lambda = - \left. \frac{\partial K\left( m^2 t, X
\right)}{\partial X}\right|_{X=m^4/2}
    - \frac{2}{\kappa^2} \dot H - \sum_i \left(1+w_i\right) \rho_{0i}
a^{-3(1+w_i)}\, .
\ee
In case of $K(\phi,X)=K(X)$ action (\ref{c1}) has the
following form:
\be
\label{x1}
S= \int d^4 x \sqrt{-g} \bigg \{
\frac{R}{2\kappa^2} + K(X) + \lambda \Big( X- U(\phi) \Big)
    + L_\mathrm{matter} \bigg \}\, ,\quad X \equiv - \frac{1}{2}\partial^\mu \phi
\partial_\mu \phi\, .
\ee
If we set $U(\phi)=\frac{m^3 \phi}{2}$, we find $\phi =m^3
t^2/2$ and therefore $U = X = m^6 t^2/2$. Then the second equation
of (\ref{c2}) determines the form of $K(X)$ as follows,
\be
\label{xx1} K(X)= - \frac{1}{\kappa ^2} (2\dot H(\sqrt{2X/m^6}) +
3H^2 (\sqrt{2X/m^6}))
    - \sum _i w_i \rho _{0i} a(\sqrt{2X/m^6})^{-3(1+w_i)}\, .
\ee
By differentiating the second equation of (\ref{c2}), we find
\be
\label{x2}
m^6 t K'(m^3 t^2/2)= - \frac{1}{\kappa^2} (2 \ddot H +
6 H \dot H) +3H \sum _i w_i (1+w_i) \rho _{0i} a^{-3(1+w_i)}\, .
\ee
Then combining (\ref{x2}) with the first equation of
(\ref{c2}), the time dependence of $\lambda$ follows:
\be
\label{xx2}
\lambda = -\bigg( \frac{2\dot H}{\kappa^2} + \sum_i
(1+w_i) \rho _{0i} a^{-3(1+w_i)} \bigg) \frac{1}{m^6 t^2} + \bigg(
\frac{2}{\kappa^2}(\ddot H + 3H \dot H) - 3H \sum_i w_i (1+w_i)
\rho _{0i}a^{-3(1+w_i)} \bigg) \frac{1}{m^6 t}\, .
\ee
Using above technique the specific examples of DE cosmology may be
reconstructed. Theory remains to be consistent but the reconstruction examples
qualitatively change. This means that the same potential which was used to
produce given cosmology leads to different cosmology in the presence of
Lagrange multiplier. Moreover, explicit realization of reconstruction scenario
turns out to be significally simpler than without constraint.

\section{FRW cosmology in $F(R)$-gravity with Lagrange multiplier}

In this section we consider $F(R)$-gravity where Lagrange
multiplier is introduced in the same way as in scalar theory of
previous section.
In the usual $F(R)$-gravity, there appears the scalar mode called scalaron,
which
often affects  the Newton law. In this section, we try to
suppress the propagation of the scalaron by imposing the constraint under the
Lagrange multiplier field. As a result, however, there seems to appear the
propagating mode in the Lagrange multiplier field, which may break the Newton
law but in somehow easier way. The solution of this question may request the
additional modification of constraint.
Another purpose of this section is the reconstruction. In usual $F(R)$-gravity,
we need to solve the complicated differential equation
\cite{capo} to realize the reconstruction program.
In this section, we show that the reconstruction can be done more easily in the
model with the Lagrange multiplier field.
The starting action is given by
\be
\label{Lag1} S = \int d^4 x \sqrt{-g} \left\{ F_1(R) - \lambda
\left( \frac{1}{2}
\partial_\mu R \partial^\mu R
+ F_2 (R) \right) \right\}\, .
\ee
Here $\lambda$ is the Lagrange multiplier field, again, which
gives a constraint
\be
\label{Lag2}
\frac{1}{2} \partial_\mu R \partial^\mu R + F_2
(R) = 0\, .
\ee
On the other hand, by the variation of the metric
$g_{\mu\nu}$, we obtain an equation corresponding to the Einstein
equation:
\be
\label{Lag3}
0 = \frac{1}{2} g_{\mu\nu} F_1(R) 
+ \frac{\lambda}{2} \partial_\mu R \partial_\nu R
+ \left( -
R_{\mu\nu} + \nabla_\mu \nabla_\nu - g_{\mu\nu} \nabla^2 \right)
\left( F_1'(R) - \lambda F_2' (R) - \nabla^\mu \left(\lambda
\nabla_\mu R \right) \right)\, .
\ee
If the Ricci curvature is
covariantly constant and the scalar curvature is a constant:
\be
\label{Lag4}
R_{\mu\nu} = \frac{R_0}{4}g_{\mu\nu}\, , \quad
R=R_0 \, ,
\ee
Eqs. (\ref{Lag2}) and (\ref{Lag3}) reduce to
\bea
\label{Lag5}
0 &=& F_2 (R_0)\, , \\
\label{Lag6} 0 &=& F_1(R_0) - \frac{1}{2} R_0 \left( F_1' (R_0) -
\lambda F_2' (R_0) \right)\, .
\eea
If Eq.(\ref{Lag5}) has a
solution, Eq. (\ref{Lag6}) can be solved with respect to the
Lagrange multiplier field:
\be
\label{Lag6b}
\lambda = \frac{ - F_1 (R_0) + R_0
F_1'(R_0)}{F_2'(R_0)}\ .
\ee
Then if $R_0$ is positive the above
solution describes de Sitter space-time which may correspond to
dark energy or inflationary epoch (for the proposal of
gravitational unification of inflation with dark energy in
modified gravity, see \cite{review}). For spatially-flat FRW
metric Eqs. (\ref{Lag2}) and $(\mu,\nu) = (0,0)$-component of
(\ref{Lag3}) have the following form:
\bea
\label{Lag8}
0 &=& - \frac{1}{2}{\dot R}^2 + F_2(R) \, ,\\
\label{Lag9} 
0 &=& - \frac{1}{2} F_1(R) 
+ 18 \lambda \left(\ddot H + 4 H \dot H \right)^2 
+ \left\{3\left( \dot H +
H^2 \right) - 3H\frac{d}{dt} \right\} \left\{ F_1'(R) - \lambda
F_2' (R) + \left(\frac{d}{dt} + 3H \right) \left( \lambda
\frac{dR}{dt} \right) \right\} \, .
\eea
When $F_2(R)>0$, Eq. (\ref{Lag8}) may be solved as
\be
\label{Lag10}
t = \int^R
\frac{dR}{\sqrt{2F_2(R)}}\, ,
\ee
which can be solved with respect
to $R$ as a function of $t$ $R=F_R(t)$. Since
\be
\label{Lag11}
R = 6 \frac{dH}{dt} + 12 H^2\, ,
\ee
one can
find the behavior of $H=\frac{\dot a}{a}$ by solving the
differential equation
\be
\label{Lag12}
6 \frac{dH}{dt} + 12 H^2
=F_R(t) \, ,
\ee
By using the obtained solution for $H=H(t)$ (and $R=F_R(t)$),
Eq.(\ref{Lag9}) becomes a differential equation for the multiplier
field $\lambda$ and we can find the behavior of $\lambda$,
$\lambda=\lambda(t)$.

Conversely when the behavior of $H(t)$ is known from the
observational data, one may reconstruct $F_2(R)$ to reproduce the
behavior of $H(t)$ by using (\ref{Lag8}). $H(t)$ gives the
behavior of $R$ as $R=R(t)$, which can be solved with respect to
$t$ as $t=t(R)$. Using (\ref{Lag8}), the explicit form of $F_2(R)$
is found to be
\be
\label{Lag14}
F_2(R) = \frac{1}{2}\left.
\left(\frac{d R}{dt}\right)^2 \right|_{t=t(R)}\, .
\ee
Note that
$F_1(R)$ can be arbitrary function.
Then the reconstruction of model can be more easily performed than that in the
usual $F(R)$-gravity.
As an explicit example, we may
consider
\be
\label{Lag15} H(t) = \frac{h_0}{t}\, ,
\ee
where $h_0$ is a constant. Then
\be
\label{LAg16} R = \frac{- 6h_0 + 12
h_0^2}{t^2}\,\quad \mbox{or} \quad t = \sqrt{\frac{- 6h_0 + 12
h_0^2}{R}}\, .
\ee
And therefore, we find
\be
\label{Lag17}
\frac{dR}{dt} = -
\frac{12\left( - h_0 + 2 h_0^2\right)}{t^3} = - \frac{2
R^{\frac{3}{2}}}{\sqrt{6 \left( - h_0 + 2 h_0^2 \right) }}\, ,
\ee
which gives
\be
\label{Lag18}
F_2(R) = \frac{R^3}{12 \left( - h_0
+ 2 h_0^2 \right) }\, .
\ee
Another example is given by
\be
\label{dS1}
R = \frac{R_-}{2} \left( 1 - \tanh \omega t \right)
+ \frac{R_+}{2} \left( 1 + \tanh \omega t \right)\, .
\ee
Here $R_\pm$ and $\omega$ are constants.
Then $t\to \pm \infty$, $R\to \pm R_\pm$ and therefore the space-time
becomes asymptotically de Sitter. One may identify the epoch of $t\to -\infty$
as inflation and $t\to +\infty$ as late acceleration.
Since
\be
\label{dS2}
\dot R = \frac{\left(R_- - R_+\right) \omega}{2\cosh^2 \omega t}
= \frac{\left(R_- - R_+\right) \omega}{2} \left( 1
 - \frac{\left( R_- + R_+ - 2R \right)^2}{\left( R_- - R_+ \right)^2} 
\right)\, ,
\ee
from Eq.(\ref{Lag14}), one gets
\be
\label{dS3}
F_2(R) = \frac{\left(R_- - R_+\right)^2 \omega^2}{8} \left( 1
  - \frac{\left( R_- + R_+ - 2R \right)^2}{\left( R_- - R_+ \right)^2}
\right)^2\, ,
\ee
Thus, the unification of early-time inflation with dark energy epoch is
possible also in constraint modified gravity.
Hence, the universe evolution only depends on the constraint
equation (\ref{Lag8}) but does not depend on $F_1(R)$. $F_1(R)$
can only affect the correction to the Newton law. In convenient
$F_1(R)$ cosmology the whole dynamics is defined by the form of
this function. With the constraint (\ref{Lag8}), $F_1(R)$ becomes
irrelevant. The cosmological dynamics is defined by the form of
$F_2(R)$. In $F(R)$-gravity, there appears the propagating mode,
which is often called scalaron and which often violates the 
Newton law. In the same way as in the
scalar-tensor theory around (\ref{LagS7}), we may show that the
scalaron does not propagate. In case of $F(R)$-gravity, however,
Eq. (\ref{Lag3}) contains the second derivative of the multiplier
field $\lambda$ although the Einstein equation (\ref{LagE}) of the
scalar theory with a constraint (\ref{LagS1}) does not contain the
derivative of the multiplier field $\lambda$ and therefore
$\lambda$ can be solved algebraically as in (\ref{LagS6}). In case
of $F(R)$-gravity, we need to solve the second order differential
equation to find $\lambda$, which might indicate that $\lambda$ could
propagate and there might appear the correction to the Newton
law. The magnitude of the correction could depend on the choice of
$F_1(R)$ and/or $F_2(R)$.

In order to investigate the Newton law, we choose $F_1(R)$ as the
Einstein one,
\be
\label{Lag19} 
F_1(R) = \frac{R}{2\kappa^2}\, ,
\ee
and introduce the matter. Then Eq. (\ref{Lag3}) has the
following form:
\be
\label{Lag20} 0 = \frac{1}{2\kappa^2}
\left(\frac{1}{2} g_{\mu\nu} R - R_{\mu\nu} \right) + \frac{1}{2}
T_{\mu\nu}
    - \left( - R_{\mu\nu} + \nabla_\mu \nabla_\nu - g_{\mu\nu} \nabla^2 \right)
\left( \lambda F_2' (R) - \nabla^\mu \left(\lambda \nabla_\mu R
\right) \right)\, .
\ee
For the solution where $\lambda=0$,
Eq.(\ref{Lag3}) reduces to the Einstein equation,
\be
\label{Lag21} 0 = \frac{1}{\kappa^2} \left(\frac{1}{2} g_{\mu\nu}
R - R_{\mu\nu} \right) + T_{\mu\nu} \, .
\ee
Here $T_{\mu\nu}$ is the matter energy-momentum tensor. In the
case without matter $T_{\mu\nu}=0$, the Schwarzschild space-time,
where $R=R_{\mu\nu}=0$ is a solution, which also satisfies the
constraint equation (\ref{Lag2}) if $F_2(0)=0$. In case with
matter $T_{\mu\nu}\neq 0$, however, the Einstein equation
(\ref{Lag21}) gives \be \label{Lag22} R = - \kappa^2 T\, . \ee

Here $T$ is the trace of the energy-momentum tensor. The constraint 
equation (\ref{Lag2}) is
rewritten to
\be
\label{Lag23}
0 = \frac{\kappa^4}{2} \partial_\mu
T \partial^\mu T + F_2 \left( - \kappa^2 T \right)\, ,
\ee
which
is not always satisfied. Hence, in the presence of the matter, the
constraint equation (\ref{Lag2}) should be modified to be
\be
\label{Lag24}
0 = \frac{1}{2} \partial_\mu R \partial^\mu R + F_2
(R) - \frac{\kappa^4}{2} \partial_\mu T \partial^\mu T - F_2
\left( - \kappa^2 T \right)\, .
\ee
This indicates that the total constrained action with matter
should be, instead of (\ref{Lag1},
\be
\label{Lag25}
S = \int d^4
x \sqrt{-g} \left[ \frac{R}{2\kappa^2} - \lambda \left\{
\frac{1}{2} \partial_\mu R \partial^\mu R + F_2 (R) -
\frac{\kappa^4}{2} \partial_\mu T \partial^\mu T - F_2 \left( -
\kappa^2 T \right) \right\} + \mathcal{L}_\mathrm{matter}
\right]\, ,
\ee
In this case, the Newton law could be easily reproduced. Here
$\mathcal{L}_\mathrm{matter}$ is the Lagrangian of the matter.
Of course, qualitatively other form of constraint 
may also solve this problem.
We also note that the form of the constraint could be correct when 
$F_1(R)$ is given by (\ref{Lag19}). 
For general $F_1(R)$, the constraint could be changed. 

Since $T$ vanishes in the vacuum as in the bulk of the universe, 
the constraint (\ref{Lag24}) reduces to 
\be
\label{Lag24b}
0 = \frac{1}{2} \partial_\mu R \partial^\mu R + F_2
(R) - F_2 \left( 0 \right)\, .
\ee
If $F_2 \left( 0 \right)=0$, the constraint (\ref{Lag24}) gives 
(\ref{Lag2}) and (\ref{Lag8}) and the cosmological evolution could 
be generated.  
Note that $F_2 \left( R \right)$ in (\ref{Lag18}) satisfies the condition 
$F_2 \left( 0 \right)=0$ but $F_2 \left( R \right)$ in (\ref{dS3}) does not. 
In case the condition $F_2 \left( 0 \right)=0$ is satisfied, there are 
two classes of solution in the constraint (\ref{Lag8}). One is a trivial 
solution $R=0$ and another corresponds to the non-trivial cosmological 
evolution given by (\ref{Lag10}). 
Near the solar systems and the galaxies, the solution could correspond 
to the trivial one $R=0$ in order to reproduce the Newton law but in the bulk 
of the universe, the solution should correspond to (\ref{Lag10}) so that the 
evolution of the universe could be generated. 
It is not so trivial to show or to deny that the two solutions could be connected 
in the intermediate region between the region near the solar systems or 
galaxies and the region of the bulk universe. 
Maybe we need more careful 
(possibly numerical) analysis, which requests a future investigation in 
this direction.

Thus, we demonstrated the possibility to describe the cosmological
dynamics, including dark energy era, in modified $F(R)$-gravity
with Lagrange constraint. It turns out that reconstruction which
produces the viable cosmology in this case is realized qualitatively easier
than in the convenient modified gravity. Moreover, to
pass the local tests may request the additional modifications of
constraint equations as is seen for emergence of Newton law
regime. This may be caused by the fact that Lagrange multiplier
propagates in such a theory. It is quite interesting observation because
modifying the form of constraint one may arrive to qualitatively different
predictions about local tests, which can make the same theory with different
constraint to be viable!

\section{ Lagrange multipliers in generic higher-order theories}

Let us discuss other application of Lagrange multiplies. In the optimization
problems, such a method
allows to find out extremal points (maxima and minima) of a
function where one or more than one constraints are present.
Specifically, Lagrange multipliers allow to calculate stationary
points of the constrained function. In other words, the method
reduces the search for stationary points of a $n$-variable
function with $k$-constraints to find out the stationary points of
free (non-constrained) function of $n+k$-variables: it introduces
a new (unknown) scalar variable, the Lagrange multiplier, for each
constraint present in the problem defining a new function (the
Lagrangian) in terms of the original function, the constraints and
the Lagrange multipliers. Up to this point, the Lagrange
multipliers have been imposed {\it a priori} to modify the
dynamics and select the form of the effective potential.
Furthermore, by integrating the multipliers, cosmological
solutions have been achieved. On the other hand, it is possible to
show that the Lagrange multipliers are constraints capable of
reducing the dynamics in higher order theories. Technically they
are anholonomic constraints being time-dependent. They give rise
to field equations which describe the dynamics of the further
degrees of freedom coming from higher order theories. This fact is
relevant to deal with such new degrees of freedom under the
standard of effective scalar fields.

With these considerations in mind, let us take into account
generic higher--order theories described by the action
\begin{equation}
\label{salv1}
    {\cal A}=\int d^4x \sqrt{-g} F(R, \Box R, \Box^2R, \ldots, \Box^kR)\,{.}
\end{equation}
The field equations are
\begin{eqnarray}
R^{\mu\nu}-\frac{1}{2}g^{\mu\nu}R&=&\frac{1}{{\cal
G}}\left\{\frac{1}{2}g^{\mu\nu}(F-{{\cal G}}R)
+(g^{\mu\lambda}
    g^{\nu\sigma}-g^{\mu\nu}g^{\lambda\sigma}){\cal
G}_{;\,\lambda\sigma}+\right.
    \nonumber \\
    & + &
    \frac{1}{2}\sum_{i=1}^k\sum_{j=1}^i(g^{\mu\nu}g^{\lambda\sigma}+g^{\mu\lambda}
    g^{\nu\sigma})(\Box^{j-i})_{;\,\sigma}\left(\Box^{i-j}\frac{\partial
F}{\partial\Box^iR}
    \right)_{;\,\lambda}+\nonumber \\
    &
    -&\left.
g^{\mu\nu}g^{\lambda\sigma}\left[(\Box^{j-1})_{;\,\sigma}\Box^{i-j}
    \frac{\partial F}{\partial\Box^iR}\right]\right\}\,{,}
    \label{salv2}
\end{eqnarray}
where
\begin{equation}\label{eq9}
    {\cal G}=\sum_{j=0}^{k}\Box^j \left(\frac{\partial
    F}{\partial\Box^jR}\right)\,{.}
\end{equation}
These are pure gravity $(2k+4)$--order field equations. Matter can
be taken into account by introducing, as above, the
energy--momentum tensor $T_{\mu\nu}$ .

In order to better discuss the role of Lagrange multipliers, let
us consider, for the moment, actions containing up to $\Box R$
terms. In this case, we have eight--order field equations which
becomes of sixth--order if the theories is linear in $\Box R$. If
we take into account FRW point--like actions, we can reduce to
the Lagrangian
\begin{equation}\label{salv4}
{\cal L}={\cal L}(a, \dot{a}, R, \dot{R}, \Box R, \dot{(\Box
R)})\,{,}
\end{equation}
by which one can deduce the Euler-Lagrange equations corresponding
to the Friedmann equations of cosmology. It is easy to show that
such cosmological equations follow from
Einstein gravity so deriving field equations from a field
action and then reducing them to cosmological equations or
reducing the field Lagrangian to a point--like Lagrangian and then
deducing the Euler--Lagrange equations gives exactly the same
results \cite{NuovoCimento}. In Eq.(\ref{salv4}), dot represents
derivative with respect to cosmic time and, as standard for
cosmological Lagrangian deduced from field theories, the
covariance is lost. The variables $R$ and $\Box R$ can be
considered independent and, by the method of Lagrange
multipliers, we can eliminate higher than one time derivatives. If
we would not consider Lagrange multipliers, the Lagrangian
(\ref{salv4}) cannot be considered canonical\cite{masud}. The action related
to Lagrangian (\ref{salv1}), up to $\Box R$ terms becomes
\begin{equation} \label{salv5}
    {\cal A}=2\pi^2\int dt\left\{
a^3F-\lambda_1\left[R+6\left(\frac{\ddot{a}}{a}+\left(\frac{\dot{a}}{a}\right)^2
+\frac{k}{a^2}\right)\right]-\lambda_2\left[\Box
R-\ddot{R}-3\,\left(\frac{\dot{a}}{a}\right)\dot{R}\right]\right\}\,{.}
\end{equation}
$\lambda_{1,2}$ are given by varying the action with respect to
$R$ and $\Box R$, that is
\begin{equation}\label{mult1}
\lambda_1=a^3\,\frac{\partial F}{\partial R}\,{,} \quad
\lambda_2=a^3\frac{\partial F}{\partial (\Box R)}\,{.}
\end{equation}
Only in this case the action results canonically defined in terms
of $R$ and $\Box R$ considered as {\it independent} variables.
After an integration by parts, the point--like Lagrangian results
\begin{equation}
    {\cal L}=6a\dot{a}^2\frac{\partial F}{\partial R}+6a^2\dot{a}\frac{d}{dt}
    \,\left(\frac{\partial F}{\partial R}\right) -a^3\dot{R}\frac{d}{dt}
    \left(\frac{\partial F}{\partial (\Box R)}\right)
    + a^3\left[F-\left(R+\frac{6k}{a^2}\right)\frac{\partial F}{\partial
R}-\Box R\,\frac{\partial F}{\partial (\Box R)}\right]
    \,{,}\label{salv6}
\end{equation}
where, clearly, the canonically conjugate variable of
configuration space are the set ${\cal Q}=\{a,R,\Box R\}$ and the
relative {\it velocities}. A remark is necessary at this point. We
can also take into account
\begin{equation} \label{salv7}
    \lambda_1=a^3\left[\frac{\partial F}{\partial R}+\Box
    \frac{\partial F}{\partial (\Box R)}\right]\,{,}
\end{equation}
as a Lagrange multiplier \label{masud}.
The Lagrangian which comes out differs
from (\ref{salv6}) just for a term vanishing on the constraint,
being
\begin{equation}
    \tilde{\cal L}={\cal L}-
    a^3\left\{R+6\left[\frac{\ddot{a}}{a}+\left(\frac{\dot{a}}{a}\right)^2
    +\frac{k}{a^2}\right]\right\}\Box\frac{\partial F}{\partial (\Box R)}\,.
\end{equation}
From this point of view, considering the point--like Lagrangian
${\cal L}$ or $\tilde{\cal L}$ is completely equivalent.

It is important to stress that Lagrange multipliers are constraints that,
after variation, give rise to further equations of motion (one for any
multipliers). In fact,  the expression that are multiplied by  Lagrange
multipliers in the action are constraints. The variation of the action
with respect to the Lagrange  multipliers gives  equations of motion of
the form "constraint equal to zero". Alternatively,  one can  solve the
Lagrange multipliers and insert them into the action. From the resulting
action one  obtains, of course,  the same constraint equations. To show this point,
let us derive the Euler--Lagrangian equations from the Lagrangian
(\ref{salv6}) that is defined on the tangent bundle ${\cal TQ}\equiv
\{a,\dot{a},R,\dot{R},\Box R, (\dot{\Box R})\}$. They can be also deduced from
the Einstein
equations (\ref{salv2}). The equation for the variables $\{a,\dot{a}\}$
gives
\begin{eqnarray}
    & &\left[R\frac{\partial F}{\partial R}+\Box R \frac{\partial
F}{\partial (\Box R)}
    -F\right]+2\left[3H^2+2\dot{H}+
    \frac{k}{a^2}\right]\frac{\partial F}{\partial R}+ \nonumber \\
    &+&2\left[\Box R-H\dot{R}\right]\frac{\partial^2 F}{\partial R^2}
    +\dot{R}\dot{(\Box R)}\frac{\partial^2 F}{\partial (\Box R)^2}+
    \left[2\Box^2R-2H\dot{(\Box R)}+\dot{R}^2\right]
    \frac{\partial^2 F}{\partial R\partial (\Box R)}+ \nonumber \\
    &+&2\dot{R}^2\frac{\partial^3 F}{\partial R^3}+2\dot{(\Box R)}^2
    \frac{\partial^3 F}{\partial R\partial (\Box R)^2}+4\dot{R}\dot{(\Box R)}
    \frac{\partial^3 F}{\partial R^2\partial (\Box R)}=0\,{.}
    \label{salv9}
\end{eqnarray}
The equation for $\{R,(\dot{R})\}$ gives
\begin{equation}\label{salv11}
    \Box\frac{\partial{ F}}{\partial (\Box R)}=0\,{.}
\end{equation}
Finally, the equation for the pair $\{\Box R,\dot{\Box R}\}$
coincides with the Lagrange multipliers
\begin{equation}\label{salv13}
    \Box R=\ddot{R}+3H\dot{R}\,{,}
\qquad
    R=-6\left(\dot{H}+2H^2+\frac{k}{a^2}\right)\,{.}
\end{equation}
    The energy condition,
that is the $(0,0)$--Einstein equation, gives
\begin{equation}\label{salv15}
    H^2\left(\frac{\partial F}{\partial R}\right)+
H\frac{d}{dt}\left(\frac{\partial F}{\partial R}\right)+
    \frac{1}{6}\left[\left(R+\frac{6k}{a^2}\right)\frac{\partial F}{\partial
    R}+\Box R \frac{\partial F}{\partial (\Box R)}-F-\dot{R}\frac{d}{dt}
    \left(\frac{\partial F}{\partial (\Box R)}\right)\right]\,{.}
\end{equation}
This derivation cleary show that dynamics can be made canonical by
Lagrange multipliers. However, considering further higher--order
$\Box^k R$ terms the process can be made iterative since for each
2-orders (i.e. $\Box$) one has another Lagrange multiplier. This
method allows to select suitable changes of variables that can be
identified once the variables $a, R$ and $\Box R$ are
disentangled \cite{NuovoCimento}. As a consequence, dynamics can be reduced and
exactly integrated \cite{defelice,prado}.

An important remark is in order at this point. If we take into account a
conformal transformation as (\ref{LagBD3}), it is easy to show that
\begin{equation}\label{salv17}
    \tilde{g}_{\mu\nu}\equiv
    \left(\frac{dF}{dR}\right)\,g_{\mu\nu}\,{,}\quad
    \phi=\sqrt{\frac{3}{2}}\ln\left(\frac{dF}{dR}\right)\,{,}
\end{equation}
for a $F(R)$-gravity and
\begin{equation}\label{mult2}
    \tilde{g}_{\mu\nu}\equiv
    \left(\frac{\partial F}{\partial R}+\Box \frac{\partial F}{\partial\Box
R}\right)\,g_{\mu\nu}\,{,}\quad
    \phi=\sqrt{\frac{3}{2}}\ln\left(\frac{\partial F}{\partial R}+\Box
\frac{\partial F}{\partial\Box R}\right) \,{,}
\end{equation}
for $F(R,\Box R)$-gravity.
It is easy to see that such transformations are related to the Lagrange
multipliers (\ref{mult1}) and (\ref{salv7}).
This means that operating a conformal transformation from the Jordan frame to
the Einstein frame,
that is disentangling the additional gravitational degrees of freedom related
to higher order
theories of gravity has the same dynamical meaning of reducing the dynamical
system by imposing Lagrange multipliers.

\section{Conclusions}

In summary, we studied the role of Lagrange multiplier constraint for DE
cosmology in scalar-tensor and quintessence theory and modified
$F(R)$-gravity.
It is demonstrated that the presence of constraint significally simplifies
the reconstruction scenario.
Moreover, the details of cosmological evolution are qualitatively changed.
   For instance, the phantom/non-phantom
transition in scalar-tensor theory is more smooth.
In modified gravity, the presence of Lagrange multiplier induces the necessity
to introduce second $F(R)$ function which plays major role for cosmology.
Again, the reconstruction program is qualitatively simplified.
It is shown that viable reconstruction may be achieved.
For instance, the examples of dark energy era and unified
   early-time inflation with late-time acceleration are worked out.
We considered the constraint of specific form which permits to change
the EoS parameter of the effective scalar fluid.
However, many other choices for constraint may be useful.
For instance, additional modification of Lagrange multiplier constraint
may help to pass the local tests for the theory which cannot pass local 
tests in its original formulation.

A general comment is in order at this point. Gravitational
theories are {\it constrained theories}. Such constraints can be
anholonomic and then can result as further equations of motion for
the related dynamical systems. We have first investigated the
possibility that introducing {\it by hand} Lagrange multipliers,
we can select suitable forms of the effective potential $V(\phi)$
and of $F(R)$-gravity. This approach leads to solvable DE dynamics for
several physically interesting models. Besides, we have considered
higher--order models. Lagrange multipliers allow, in this case, to
make the theory canonical, that is they allow to disentangle the
degrees of freedom of the problem and then reduce the dynamics. In
some examples that we have worked out, it is possible to achieve
exact solutions thanks to the multipliers that $i)$ result as new
equations of motion $ii)$ allow suitable change of variables
which, identifying cyclic variables, are related to constant of
motion and allow to reduce and integrate the dynamics. The two
methods (i.e. imposing the multipliers a priori or using them to
reduce dynamics and make it canonical) are effectively equivalent
and show the possibility to develop a new approach to alternative
theories of gravity.

An important consideration is in order for conformal
transformations. In the Einstein frame, gravitational degrees of
freedom and scalar field are well separated. Imposing Lagrange
multipliers means to search for suitable forms of the scalar field
potential that allow to integrate dynamics. In this case, the
Lagrange multipliers act as a "selection rule" on the dynamics and
give fixed stationary points. On the other hand, in the Jordan
frame, dynamics is not "canonical" since gravitational degrees of
freedom and/or scalar fields are not disentangled. Lagrange
multipliers, as said, make dynamics canonical. It is interesting
to see that the form of conformal transformation can be related to
the Lagrange multipliers (they have the same forms, see Eqs.
(\ref{mult1}), (\ref{salv7}), (\ref{salv17}), and
({\ref{mult2})). In some sense we can state that imposing
Lagrange multipliers in the Jordan frame and performing conformal
transformations to the Einstein frame are the same operation. Also
the reverse statement holds: given a Lagrangian in the Einstein
frame endowed with Lagrange multipliers means, under conformal
transformation to the Jordan frame, to take into account higher
order or non-minimally coupled theories that are, in any case,
canonical. This question will be investigated elsewhere.

\section*{Acknowledgments}

This work has been suported by INFN-MEC collaboration project. The work by S.N.
is supported in part by Global COE Program of
Nagoya University provided by the Japan Society for the Promotion
of Science (G07) and that by S.D.O. is supported in part by MEC (Spain) project
FIS2006-02842 and LRSS project 3558.2010.2.

\end{document}